# Stray light in 3D porous nanostructures of single crystalline copper film

Yu-Seong Seo[1], Teawoo Ha[2], Ji Hee Yoo[3], Su Jae Kim[4], Yousil Lee[4], Seungje Kim[1], Young-Hoon Kim[5], SeungNam Cha[1], Young-Min Kim[5], Se-Young Jeong[6,7,8*], and Jungseek Hwang[1*]

[1]*Department of Physics, Sungkyunkwan University (SKKU), Suwon 16419, Republic of Korea*
[2]*Center for Integrated Nanostructure Physics (CINAP) & Center for 2D Quantum Heterostructures, Institute for Basic Science (IBS), Suwon 16419, Republic of Korea*
[3]*Department of Cogno-Mechatronics Engineering, Pusan National University, Busan 46241, Republic of Korea*
[4]*Crystal Bank Research Institute, Pusan National University, Busan 46241, Republic of Korea*
[5]*Center for Integrated Nanostructure Physics (CINAP), Institute for Basic Science, Suwon 16419, Republic of Korea*
[6]*Gordon Center for Medical Imaging, Department of Radiology, Massachusetts General Hospital and Harvard Medical School, Boston, MA, 02114, USA.*
[7]*Department of Optics and Mechatronics Engineering, Pusan National University Busan 46241, Republic of Korea, Department of Physics*
[8]*Korea Advanced Institute of Science and Technology (KAIST), Daejeon 34141, Korea*

Corresponding e-mails: syjeong@pusan.ac.kr or jungseek@skku.edu

## Abstract

In the design of optical devices and components, geometric structures and optical properties of materials, such as absorption, refraction, reflection, diffraction, scattering, and trapping, have been utilized. Finding the ideal material with certain optical and geometric characteristics is essential for a customized application. Here, we fabricated unoxidizable achromatic copper films (ACFs) on $Al_2O_3$ substrates utilizing an atomic sputtering epitaxy apparatus. ACFs are made up of two regions vertically: a comparatively flat layer region and a three-dimensional (3D) porous nanostructured region on top of the flat region. The measured specular reflectance displayed low-pass filter behaviour with a sharp cutoff frequency in the infrared spectrum. Furthermore, the measured diffusive reflectance spectra showed light-trapping behaviour in the spectral region above the cutoff frequency, where there are no known absorption mechanisms, such as phonons and interband transitions. A focused ion beam scanning electron microscope was utilized to study the thin film's nanostructured region through 3D tomographic analysis in order to comprehend the phenomena that were observed. This work will shed fresh light on the design and optimization of optical filters and light-trapping employing porous nanostructured metallic thin films.

## 1. Introduction

Optical components and devices have been constructed by utilizing the optical characteristics of materials, such as absorption, refraction, and reflection, as well as their geometric structures, such as diffraction, interference, and scattering.[1,2] Thus far, various optical devices, such as lenses, mirrors, polarizers, optical filters, optical fibers, metamaterials, and photonic crystals have been developed.[3–7] Polarizers have been developed using a selective absorption process; infrared filters use the scattering and absorption effects; optical fibers use the refraction phenomenon; and metamaterials and photonic crystals use periodic geometric structures.[3,8–11] In recent years, numerous novel optical devices have been developed based on geometric structures [12–20] rather than the intrinsic optical properties of currently available materials because their intrinsic optical properties have been studied and utilized. Therefore, recent studies have focused on the development of new optical devices using geometric structures. For example, various colors of copper oxide/copper have been achieved by accurately controlling the thickness of the copper oxide layer. [21]

An intriguing phenomenon that can be produced with geometric constructions is light trapping. Geometric structures for light trapping have been intensively researched and employed in scientific and technical domains because they can improve the effectiveness and performance of optical systems.[22–24] In solar cells, light-trapping structures are used to enhance the absorption of solar radiation, resulting in increased energy conversion efficiency.[25,26] In addition, light-trapping structures can be used to improve the brightness and contrast of displays such as liquid crystal and organic light-emitting diode displays.[27,28] Furthermore, light-trapping structures have the potential to improve the sensitivity and selectivity of photodetectors, enabling better sensing and imaging performances.[29–32] In optical sensing and spectroscopy, light-trapping structures can be used to improve detection by multiple interactions with light and matter within a small volume. Overall, light-trapping structures represent a promising approach for enhancing the performance of optical devices and technologies. Therefore, generating photonic structures that trap light is a very important issue. The photonic structures are intimately related to the geometrical structures of materials, which enhance light-matter interactions.

In this study, we focused on generating and characterizing geometric (or photonic) structures with a porous nanostructure of a single-crystalline Cu film using a thin-film growth method known as atomic sputtering epitaxy (ASE),[33,34] to develop light-trapping structures. The geometric structures are single-crystalline achromatic Cu films (ACF) grown on an $Al_2O_3$ substrate (see the Methods section). ACF has already been verified as a groundbreaking fabrication method for single-crystalline and oxidation-resistant Cu thin films through X-ray diffraction and transmission electron microscopy (TEM).[35] Here we first measured the specular reflectance spectra of the ACF samples over a wide spectral range from far-infrared to visible. The measured reflectance spectra exhibited low-pass filter behavior with a sharp cutoff frequency. The cutoff frequency is closely related to the characteristic length scale of the geometrical surface irregularity of a thin-film sample. Furthermore, we measured the diffusive reflectance of the ACF samples using an integrating sphere over a wide spectral range from mid-infrared to visible. Interestingly, we found that some light did not come out of the ACF sample, even though there were no absorption processes in the spectral region of interest. This phenomenon was the light-trapping effect of the cavity structure in the porous nanostructure of the single-crystalline Cu sample. To determine the characteristic length scale for the cutoff frequency and understand the light-trapping effect, we investigated the geometric structure of the Cu film using focused ion beam scanning electron microscopy (FIB-SEM), scanning electron microscopy (SEM), and atomic force microscopy (AFM). We were able to indirectly identify the characteristic length scale related to the cutoff frequency and the cavity structure for the light trapping from the FIB-SEM and AMF images.

## 2. Results and Discussion

Figure 1a shows optical images of eight ACF samples on 2-inch $Al_2O_3$ wafers and four representative SEM images. The SEM images demonstrate that the achromaticity change is directly related to the systematic change in geometric features of the ACFs. There seems to be an approximately positive correlation between the open-pore volume fraction and achromaticity (see also Figure 3 and Figure S3 in Supporting Information). The color of the sample surface changed from brown (Br) to black (B) depending on the growth conditions (see the Methods section). The achromaticity of the sample roughly increases from sample Br to sample B, as indicated with an arrow. The achromaticity and lightness of the achromaticity of samples can be approximately tuned by experimental conditions such as working power, pressure, and the rotational speed of the substrate [35]. We also discriminated the samples according to their color (RGB color system) and acronyms. The acronyms are brown (Br), light gray 1 (LG1), light gray 2 (LG2), gray 1 (G1), gray 2 (G2), dark gray 1 (DG1), dark gray 2 (DG2), and black (B). From now on, we denote the samples with the acronyms shown in Figure 1a. Figure 1b shows the schematic optical path in the geometric structure of the film. The incident beam can be specularly reflected, scattered, and trapped in the structure. Figure 1c shows the specular and diffusive reflectance spectra of the brown (Br) sample. The specularly reflected, scattered, and trapped portions of the incident beam are assigned in the figure. Figure 1d shows the measured specular reflectance spectra of the prepared nine samples including a single-crystalline Cu thin-film (SCCF) in a wide spectral range from 6.2 meV to 3.1 eV. The specular reflectance spectra ($R_S(\omega)$) show clear cutoff frequencies, where the reflectance shows an abrupt drop on the high wavenumber side. Below the cutoff frequency, the reflectance was higher than 95%, and above this frequency, it exhibited a sharp drop. Therefore, the film could be used as a reflection-type low-pass filter. Because the film consists of only single-crystalline Cu,[35] there are no possible absorption mechanisms in the spectral region from the cutoff frequency to the Cu plasma edge (approximately 2.1 eV). Therefore, the cutoff reflectance is associated with the 3D porous nanostructure of the film sample. In this case, the geometric effect on the reflectance is scattering. The cutoff frequency obtained from the filtering behavior provides the corresponding characteristic length scale for the porous 3D nanostructure.[35]

Furthermore, the measured diffuse reflectance spectra ($R_D(\omega)$) of the nine samples, along with the corresponding specular reflectance spectra, are shown in Figure 1e. Unexpectedly, the diffusive reflectance spectra of the samples with porous nanostructures show a strong absorption-like feature (or a pronounced dip) in a region between the cutout frequency and the plasma edge of Cu (~ 2.1 eV). The sample with a lower cutoff frequency exhibited a lower diffusive reflectance in the region. A diffusive reflectance lower than that of the SCCF sample in this region can be understood by accounting for trapped light in cavities of the 3D porous nanostructure (Figure 3) as depicted in Figure 1b. This type of fictitious absorption can be called geometric absorption because its origin is intimately associated with light trapping in the 3D porous nanostructures of the samples.

The characteristic length of the surface roughness of a sample that causes light scattering can be estimated from the cutoff frequency in either the reflectance or transmittance spectrum.[36] In general, the cutoff frequency can be the frequency at which either transmission or reflectance decreases to 50% throughput in a low-pass filter. In the reflectance spectrum, the cutoff frequency was closely related to the characteristic length of the surface roughness of the sample.

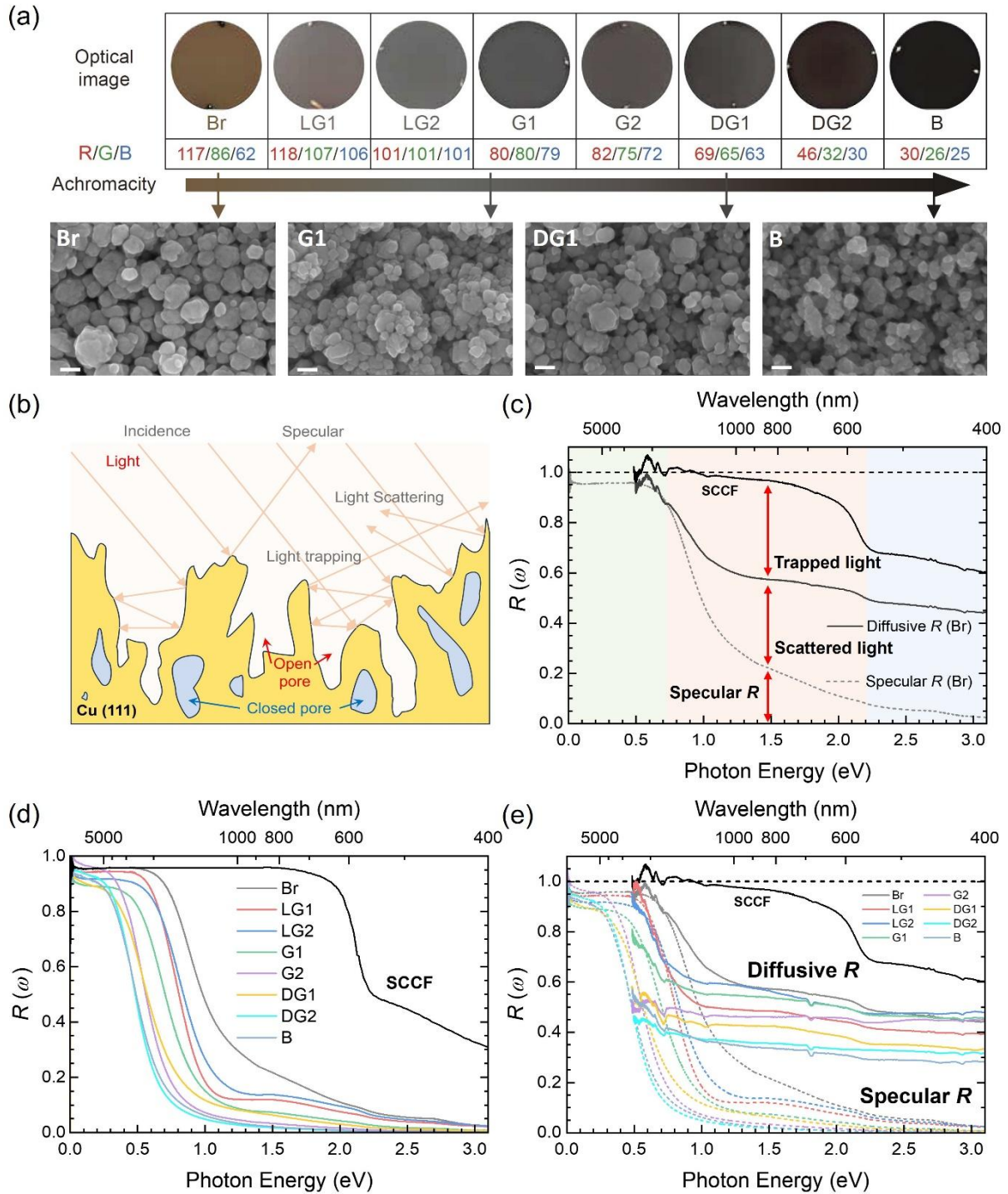

**Figure 1**. Single-crystalline copper film with porous nanostructures. a) 2-inch wafer-size thin film samples in various colors from brown to black, an arrow to show the achromaticity of samples, and SEM images of four representative samples. All the scale bars indicate 100 nm. b) Schematic diagram of beam paths of specularly reflected, scattered, and trapped lights in the 3D porous nanostructured region. c) Diffusive and specular reflectance spectra of the Br sample.

The specularly reflected, scattered, and trapped portions are assigned at 1.5 eV. d) Measured specular reflectance spectra of our nine samples e) Diffusive reflectance spectra of our samples and a 40 nm-thick single-crystalline Cu film (SCCF), along with the corresponding specular reflectance spectra for comparison.

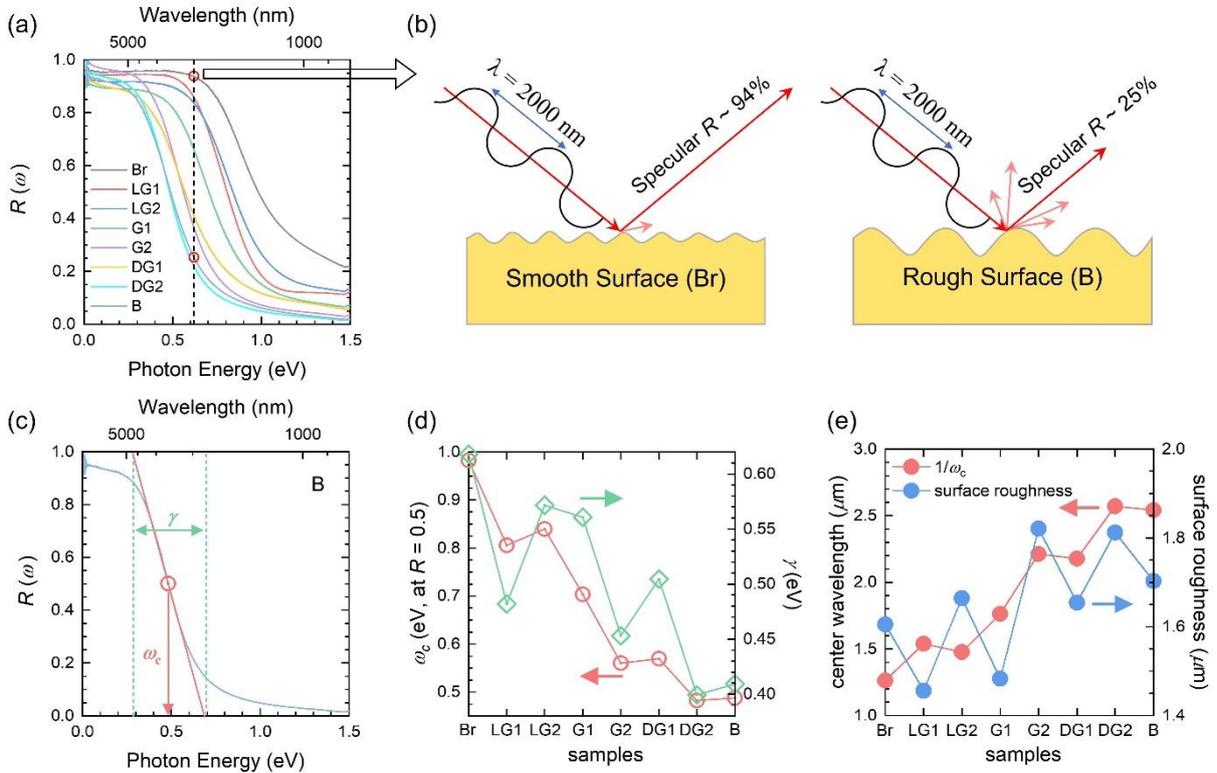

**Figure 2.** Reflectance-type low-pass filter. a) Measured specular reflectance spectra of our eight thin film samples near the cutoff frequency. b) Schematic diagrams of specular reflectance for the Br (smooth surface) and B (rough surface) samples. c) Schematic diagram of beam path depending on the relation wavelength and surface roughness average. d) Center frequencies and widths of all eight samples. e) Center wavelengths and surface roughnesses of eight samples. The surface roughness was obtained from the AFM image (see Supporting Information)

Figure 2a shows the measured specular reflectance spectra of the eight thin-film samples near the cut-off frequency below 1.5 eV. Figure 2b shows a schematic diagram of specular and scattered beams, which depend on the average surface roughness. The measured specular reflectance exhibited low pass filtering behavior. To quantify the cutoff in the reflectance, we obtained the center (or cutoff) frequency ($\omega_c$) and width ($\gamma$) of the cutoff using the method described in Figure 2c. The center frequency was estimated from the frequency at which the reflectance was 0.5 on the cutoff. The width was estimated from the frequency difference between the two crossing points on horizontal lines at $R = 0.0$ and $1.0$ with the extrapolated straight line having the same slope as the cutoff. The obtained center frequencies and widths of all eight samples are shown in Figure 2d. The center frequencies are in the mid-infrared region from 0.5 to 1.0 eV. The center frequency is approximately proportional to the width. The center frequency can be converted to the center wavelength, $\lambda_c$ (Figure 2e), which ranged from 1.25 to 2.50 $\mu$m and was closely related to the surface roughness of the porous 3D nanostructure of the thin-film surface. To obtain information on the surface roughness, AFM was performed on all eight samples (see Supporting Information). The average period of the surface roughness obtained from the power spectral density (PSD) of the AFM image correlates with the center wavelength, as shown in Figure 2e. However, the sample-dependent change is not sufficiently large enough compared to that of the center wavelength.

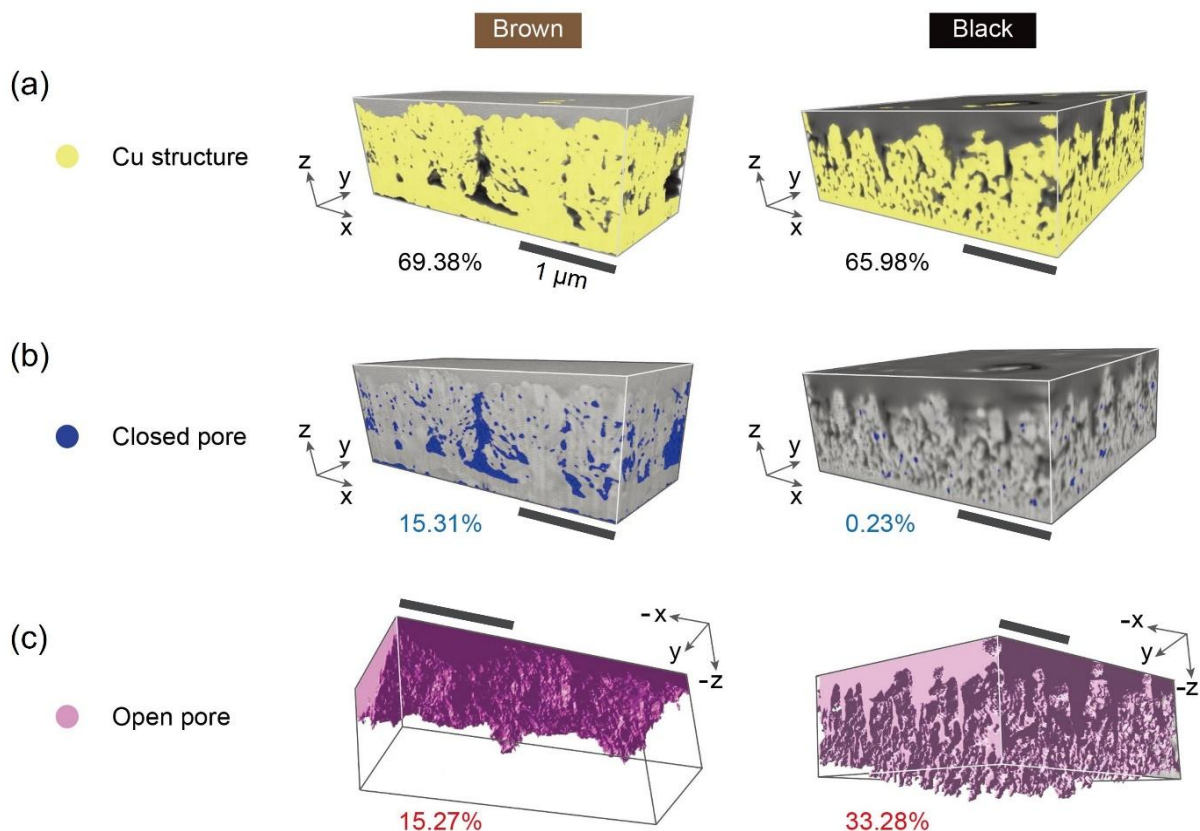

**Figure 3.** Tomography images of 3D porous nanostructured region: Cu structure a), closed pores b), and open pores c) of brown (left columns) and black samples (right columns).

The porous 3D nanostructured regions of the thin films were investigated using a spatially resolvable experimental technique, FIB-SEM (see the Methods section). The resulting 3D images of two representative samples (Br and B) are shown in Figure 3. For both samples, the 3D nanostructures of Cu in the entire measured volume and the volume ratio occupied by Cu with respect to the entire volume are shown in Figure 3a, while the closed and open pores and their volume ratios with respect to the entire volume are shown in Figures 3b and 3c, respectively. The Cu volume ratios of the two samples were comparable. However, the volume ratios of closed and open pores were significantly different between the two samples. The closed pore volume ratio of the B sample was 0.23%, which was negligibly small compared to that of the Br sample (15.31%), whereas its open pore volume ratio was 33.28%, which was twice that of the Br sample (15.27%). Figure 3c shows that the open pores of the Br sample were relatively unconnected to the lower part near the substrate, whereas those of the B sample were spread evenly over the entire Cu area and connected to the deep area near the substrate. The overall 3D nanostructure of the open pores was closely associated with the cutoff frequency and light-trapping phenomena occurring in the 3D porous Cu thin film. The average surface roughness of the Cu nanostructure (Figure 3a) is closely related to the scattering of incident light, resulting in a cutoff (or center) frequency for the low-pass filter (see Figure 1c). The incident light is trapped in cavities, each of which may consist of many open pores. Therefore, trapped light does not appear in the measured diffusive reflectance, resulting in an absorption-like feature observed between the cutoff frequency and the Cu plasma edge in the diffusive reflectance of the samples (see Figures 1c and 1e). This light trapping may influence the emissivity of 3D porous Cu films. The more trapped light, the higher the emissivity.

The ACF samples with 3D porous nanostructures may have two functions: light scattering and trapping. Both functions directly influence depend on the achromaticity of the samples. The achromaticity of the sample is intimately associated with its 3D porous nanostructure.[35] The emissivity ($\varepsilon(\omega)$) can be obtained from the measured diffusive reflectance ($R_D(\omega)$), that is, $\varepsilon(\omega)$ = 1 - $R_D(\omega)$. In general, the emissivity of a surface depends not only on the surface material but also on the surface geometric structures. Particularly, as we discussed previously, the surface geometric structure influences the diffusive reflectivity of the surface of the material. Therefore, the 3D porous nanostructure of the ACF sample affected surface emissivity.

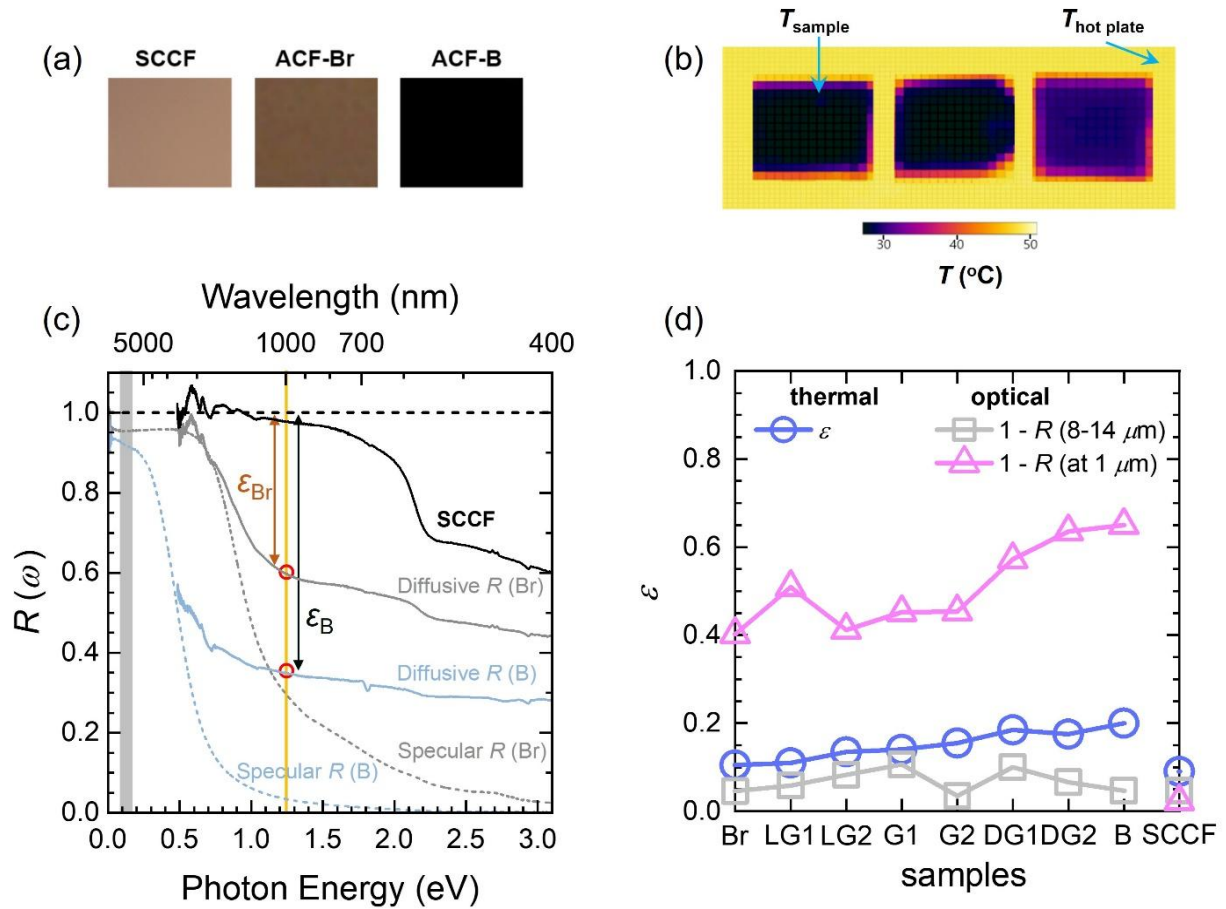

**Figure 4.** a) Optical images of SCCF and two (Br and B) achromatic copper film (ACF) samples. b) Thermal images of SCCF and the two ACFs on a hotplate obtained using an infrared thermal imaging camera. c) Specular and diffusive reflectance spectra for SCCF and the two ACF samples. The definition of emissivity ($\varepsilon$) at a given wavenumber is depicted, i.e. $\varepsilon(\omega) = 1 - R_D(\omega)$. d) Measured emissivity data of all eight samples in the infrared region (8–14 μm) using an infrared thermal camera, as well as the emissivity data obtained from the diffusive reflectance spectra at a mid-infrared frequency of 0.11 eV and at a near-infrared frequency of 1.2 eV.

Figure 4a shows the optical images of three samples: a flat SCCF and two ACF samples (Br and B). Figure 4b shows the thermal images of the three samples, along with a color bar scaled with the temperature. We determined the emissivity of each sample in the mid-infrared region using an infrared thermal-imaging camera as follows: The sample temperature was measured on a hotplate at the set temperature at which thermal equilibrium was reached. By adjusting the emissivity until the correct temperature was obtained from the sample surface, we determined the emissivities of all nine samples, including the SCCF sample. Note that the gray vertical band shows the measured spectral region of the infrared camera. We also obtained the emissivity values at two different frequencies (0.11 and 1.2 eV) obtained from the measured diffusive reflectance, as depicted in the Figure 4c. Figure 4c shows the diffusive reflectance spectra of B and Br samples, along with specular ones. $\varepsilon_B$ and $\varepsilon_{Br}$ are the emissivities of black (B) and brown (Br) samples, respectively, at 1.2 eV. In Figure 4d, the emissivity in the region between the cutoff frequency and the plasma edge of Cu increases as the achromaticity increases because the light-trapping effect became stronger owing to the 3D porous nanostructure in the film. The emissivity was also approximately proportional to the center wavelength (or surface roughness) (Figure 2e). Images in Figure 4b demonstrate the thermal shielding or anti-infrared (IR) effects of ACFs, which can be utilized for stealth (or thermal camouflage) technologies in military surveillance.[37] Additionally, ACFs' high emissivity can be used in IR cloak technologies. [38,39]

Modeling and classifying the ACFs as photonic structures might be difficult due to their stochastic geometric patterns. Nevertheless, the achromaticity and lightness of the achromaticity of the ACFs can be roughly controlled by means of experimental control factors [35]. The achromaticity and lightness of the achromaticity of ACFs are intimately correlated with two geometrical structures: the pore size distribution and the three-dimensional connectivity [35]. The light-scattering and light-trapping effects are intimately associated with the achromaticity and lightness of the ACFs, the photonic structures. As a result, light scattering and trapping can be roughly tuned experimentally.

## 3. Conclusion

The effects of 3D porous metallic nanostructures on light propagation were investigated using a 3D porous nanostructure of a single-crystalline Cu film. The presence of open pores within the 3D porous nanostructure caused light scattering and trapping. The dimensions of the surface roughness and cavity for scattering and trapping are expected to be on the micrometer scale, which is much larger than the dimensions of each open pore. Therefore, the roughness and cavities can be observed on a global scale of a few decades or hundreds of nanometers. The open pores, which are spread over the 3D nanostructured region, form global structures with micron scales, which are cavities and rough surfaces. Rough surfaces scatter incident light beams, resulting in a reduction of the specular reflectance above the cutoff frequency, whereas the cavity traps the light beam, resulting in absorption-like features in the diffusive reflectance spectrum and an increase in surface emissivity. Overall, our results provide valuable insights into the effects of 3D porous metallic nanostructures on light propagation, particularly concerning the light scattering (or cut-off frequency) and light trapping (or high emissivity) phenomena. These findings contribute to the understanding of light scattering and trapping phenomena in thin films with 3D porous metallic nanostructures and may offer new avenues for designing and optimizing optical devices with enhanced performance using 3D porous metallic nanostructures.

## 4. Experimental Section/Methods

*Sample preparation and characterization*: To grow ACFs, we employed a two-step growth method using ASE. At the first step, SCCF deposition was conducted on a 2-inch sapphire (0001) substrate by coherent sputtering up to a thickness of approximately 100 nm, followed by the second step, incoherent sputtering. At the second step, a 3D porous nanostructured region with a thickness of ~ 700 nm was deposited on top of the flat layer region with a thickness of 100 nm. The porous nanostructures that maintained the crystalline structure of the Cu lattice were fabricated by making minor adjustments to the coherent ASE conditions. This involved actions such as introducing friction to the ASE's rotating device or modifying the rotational speed. During ASE growth, the deposition of individual atoms led to the formation of single-crystalline porous nanostructures aligned along the (111) direction. The ASE conditions for SCCF growth can be slightly altered by slowly rocking the growth stage, which, in turn, enables the formation of porous Cu nanostructures on the surface. Incoherent growth occurred while still maintaining epitaxial growth conditions for each nanostructure. We used a 99.99%-pure Cu target in RF sputtering to grow ACFs of different colors on sapphire (0001) substrates. The achromatic intensity, representing the lightness of these ACFs, is highly sensitive to the growth conditions, particularly revolutions per minute (RPM) at lower working pressures. Initially, the RF sputtering pressure was maintained at $2.3 \times 10^{-3}$ Pa. To adjust the achromaticity and lightness according to the growth conditions, we controlled the working pressure, ranging from $7.2 \times 10^{-1}$ to 1.6 Pa, by introducing 99.999% argon gas. Most nanostructure films grown at a high working pressure of 1.6 Pa exhibit an achromatic or dark color (with an average of RGB indices < 45), although the darkest black Cu was obtained at 0.72 Pa with 30 RPM and 40 W power.

We varied the RF power and rotation rate within the ranges of 20–60 W and 0–60 RPM, respectively. The substrate temperature remained constant at 190 °C during the deposition process, and the film thickness was controlled by adjusting the deposition time. The substrate holder was positioned 10 cm above a single-crystal Cu target diameter of 25.4 mm. The final film thickness was determined through AFM.

*Optical measurement*: The specular reflectance spectra of the prepared samples were measured using a commercial Fourier transform infrared spectrometer (Bruker Vertex 80v, Germany), which covers a wide spectral range from far infrared to ultraviolet (6.2 meV–5.0 eV). Furthermore, we measured the diffusive reflectance spectra using a monochromatic spectrometer equipped with an integrating sphere (Jasco V-670, Japan). The relative reflectance spectra were obtained relative to a single-crystalline Cu film, which served as a reference.

*Focused ion beam–scanning electron microscope (FIB-SEM)*: Cross-beam FIB-SEM is a dual-beam instrument that combines SEM and FIB into a single instrument. A Carl Zeiss Crossbeam FIB-SEM facility was used to observe the ACF samples. During the focused ion beam process, a gallium (Ga) ion beam was configured at 30 kV and 100 pA. The thickness of each serial-section slice was precisely controlled at 7 nm. For SEM imaging, the operational parameters were set at 2 kV and 300 pA while maintaining an image pixel size of 3.489 nm.

Continuous tomography utilizing a FIB involves the stepwise processing of a sample by cutting it into thin slices with a specific thickness. The Atlas solution (ZEISS, Germany) was used to monitor the exact slice thickness during this process. Subsequently, high-resolution scanning electron microscopy was used to capture images of each individual slice. This iterative process is replicated tens to thousands of times to generate a comprehensive series of images. These collected images formed the foundation for the subsequent 3D reconstruction.

*Dragonfly software–3D reconstruction analysis*: 3D reconstruction was performed using the Dragonfly (ORS) software. This software facilitated the conversion of the tomography results into 3D data, which were subsequently processed to generate cross-sectional views in the *yz* (side) and *xy* (top) directions through *zx* plane reconstruction. Segmentation of the 3D data generated in this manner aids in comprehending the shape of the material and provides detailed insights into its structural characteristics.

**Supporting Information**

Supporting Information is available from the Wiley Online Library or from the author.

**Acknowledgements**

This work was supported by the National Research Foundation of Korea (NRF-2021R1A2C101109811 & NRF-2022R1I1A1A01068619), the Samsung Research Funding &

Incubation Center of Samsung Electronics under Project Number SRFC-MA2202-02, the Commercialization Promotion Agency for R&D Outcomes (COMPA) funded by the Ministry of Science and ICT(MSIT) (2022RMD-S08), and the BrainLink program funded by the Ministry of Science and ICT through the NRFK (2022H1D3A3A01077468). This work was partially supported by Institute for Basic Science (IBS-R036-D1). This research was supported by the SungKyunKwan University and the BK21 FOUR (Graduate School Innovation) funded by the Ministry of Education (MOE, Korea) and National Research Foundation of Korea (NRF).

**Author Contributions**

Y.-S.S., T.H., and J.H.Y. contributed equally to this study. S.J.K. and Y.L. performed the Cu thin-film growth, XRD, and EBSD measurements. S. K. and S. C. performed the AFM measurements. Y.H.K. and Y.-M.K. performed TEM measurements and analyses. Y.-S.S. and J.H. performed IR and optical measurements, respectively. S.-Y.J. and J.H. conceived and supervised the study. Y.-S.S., T.H., J.H.Y., J.H., and S. Y. J. wrote the manuscript. All authors participated in the manuscript review.